\documentclass{sigchi}

\toappear{\scriptsize Permission to make digital or hard copies of part or all of this work for personal or classroom use is granted without fee provided that copies are not made or distributed for profit or commercial advantage and that copies bear this notice and the full citation on the first page. Copyrights for third-party components of this work must be honored. For all other uses, contact the owner/author(s). Copyright is held by the author/owner(s). \\
{\emph{CHI 2016}}, May 7--12, 2016, San Jose, California, USA. \\
ACM 978-1-4503-3362-7/16/05. \\
http://dx.doi.org/10.1145/2858036.2858573}

\usepackage{balance}  % to better equalize the last page
\usepackage{graphics} % for EPS, load graphicx instead 
\usepackage[T1]{fontenc}
\usepackage{txfonts}
\usepackage{times}    % comment if you want LaTeX's default font
\usepackage[pdftex]{hyperref}
\usepackage{color}
\usepackage{textcomp}
\usepackage{booktabs}
\usepackage{ccicons}
\usepackage{todonotes}
\usepackage[utf8]{inputenc}

\makeatletter
\def\url@leostyle{%
  \@ifundefined{selectfont}{\def\UrlFont{\sf}}{\def\UrlFont{\small\bf\ttfamily}}}
\makeatother
\def\pprw{8.5in}
\def\pprh{11in}

\definecolor{linkColor}{RGB}{6,125,233}
\hypersetup{%
  pdftitle={Journeys and Notes: Designing Social Computing for Non-Places},
  pdfauthor={LaTeX},
  pdfkeywords={SIGCHI, proceedings, archival format},
  bookmarksnumbered,
  pdfstartview={FitH},
  colorlinks,
  citecolor=black,
  filecolor=black,
  linkcolor=black,
  urlcolor=linkColor,
  breaklinks=true,
}

\begin{document}

\CopyrightYear{2016}
%\setcopyright{rightsretained}
\conferenceinfo{CHI'16}{May 07-12, 2016, San Jose, CA, USA}
%\isbn{978-1-4503-3362-7/16/05}
%\doi{http://dx.doi.org/10.1145/2858036.2858573}

\title{Journeys \& Notes:\\Designing Social Computing for Non-Places}

\numberofauthors{3}
\author{%
  \alignauthor{Justin Cranshaw\\
    \affaddr{Microsoft Research}\\
    \affaddr{Redmond, WA, USA}\\
    \email{justincr@microsoft.com}}\\
  \alignauthor{Andrés Monroy-Hernández\\
    \affaddr{Microsoft Research}\\
    \affaddr{Redmond, WA, USA}\\
    \email{amh@microsoft.com}}\\
  \alignauthor{S. A. Needham\\
%    \affaddr{\ }\\
    \affaddr{California, USA}\\
    \email{needhamsa@hotmail.com}}\\
}

\maketitle

\begin{abstract}
In this work we present a mobile application we designed and engineered to enable people to log their travels near and far, leave notes behind, and build a community around spaces in between destinations. Our design explores new ground for location-based social computing systems, identifying opportunities where these systems can foster the growth of on-line communities rooted at \emph{non-places}.  In our work we develop, explore, and evaluate several innovative features designed around four usage scenarios: daily commuting, long-distance traveling, quantified traveling, and journaling. We present the results of two small-scale user studies, and one large-scale, world-wide deployment, synthesizing the results as potential opportunities and lessons learned in designing social computing for non-places.
\end{abstract}


\category{H.5.m.}{Information Interfaces and Presentation
  (e.g. HCI)}{Miscellaneous} 
  %\category{See\url{http://acm.org/about/class/1998/} for the full list of ACM classifiers. This section is required.}
  {}{}

\section{Introduction}
% Lots of reasons why ppl use LBS: practical and selfpresentation (Linquist), but all of these systems are about fixed points, but there's more to our interaction than our fixed points.
%Non-places as a critique of modernity.
\begin{figure}[t!]
\centering
  \includegraphics[width=0.49\columnwidth]{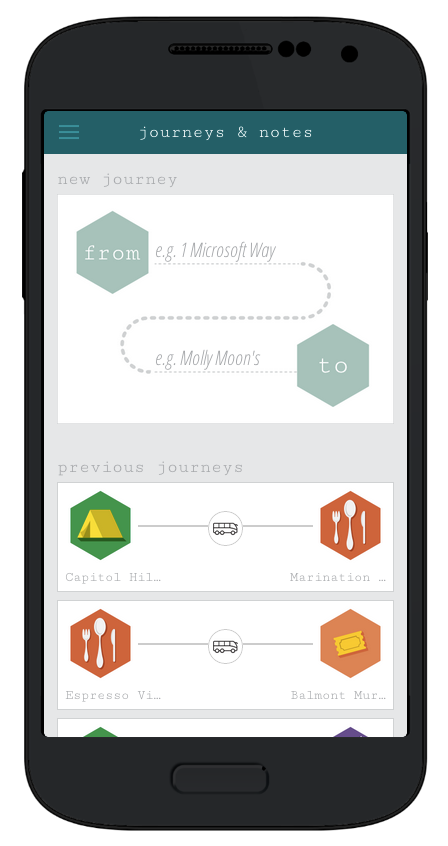}
  \includegraphics[width=0.49\columnwidth]{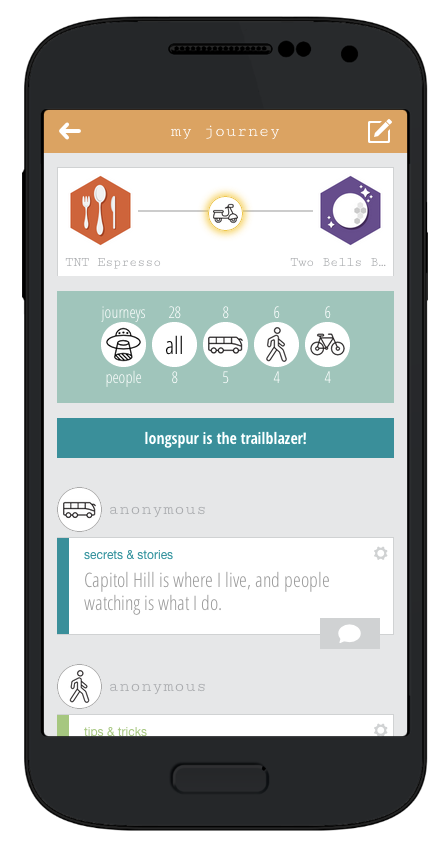}
  \caption{(Left) The \emph{Home} screen. Here users can review their history, add a new journey, and check-in to a previous journey. (Right) The \emph{Journey} screen. Once a user has checked-in to a journey, they can see stats about past visitors and interact with notes that others have left behind. }~\label{fig:mainscreens}
\end{figure}

%%%%%%%%%%%%%%%%%%%%%%%%% BEGIN INTRO %%%%%%%%%%%%%%%%%%%

The distinction between \emph{space} and \emph{place} has long provided an important theoretical framework for guiding the critical study and design of social computing systems, mobile location-based services, and online communities \cite{Harrison:1996,Dourish:2006}. Spaces are best thought of as  purely abstract---they are bounded or unbounded geographic areas or domains without any specific cultural significance. Places, on the other hand, are spaces that are imbued with a history, human social context, and cultural meaning \cite{entrikin1991betweenness, lefebvre1991production, tuan1977space, bachelard1994poetics, agnew2011space}. By understanding the processes that turn a space into a place, one is empowered with conceptual tools for understanding how to develop successful online communities and collaborative systems. 

However, it is also important to understand the characteristics of \emph{non-places} in the physical world: those physical spaces ``which cannot be defined as relational, or historical, or concerned with identity \cite{auge_non-places:_2009}." In his 1995 treatise on the subject, anthropologist Marc Aug\'{e} argues that the cultural and technological forces of modernity are pushing us to spend increasingly more of our time in non-places, describing a world where ``people are born in the clinic and die in a hospital, where transit points and temporary abodes are proliferating under luxurious or inhuman conditions (hotel chains and squats, holiday clubs and refugee camps, shanty towns threatened with demolition or doomed to festering longevity); where a dense network of means of transport which are also inhabited spaces is developing; where the habitu\'{e} of supermarkets, slot machines, and credit cards communicates wordlessly, through gestures, with an abstract, unmediated commerce \cite{auge_non-places:_2009}." 

By understanding non-places and the impact technology has on their production, we can hope future technologies will mitigate or reverse the societal trends that Aug\'{e} describes. This work aims to take a step in that direction.

We are motivated by the question of how location-based social computing can enhance the meaning, symbolization, and cultural relations experienced at physical non-places. We ask, how can place-making efforts in the digital world augment the experience at physical non-places? 

To examine these questions through design, we draw our focus on what Aug\'{e} calls the ``archetype of non-place,'' the spaces occupied by the traveller \cite{auge_non-places:_2009}. As Aug\'{e} describes ``the plurality of places'' that the traveller passes through creates a ``discontinuity between the spectator-traveller and the landscape he is contemplating or rushing through,'' preventing the him (or her) from perceiving this landscape as a place, and ``from being fully present in it \cite{auge_non-places:_2009}.'' Journeys taken by travellers often start and end at non-places (for example at bus terminals, airports, or subway stations), and are played out on the place-less fields of modern transport-infrastructure (in cars stuck in highway traffic, in cramped airplanes, or packed train cars), touching the daily lives of billions of individuals.

%By focusing our attention on designing for the non-place traveller, we transform the task of designing for non-places to one that is more concrete that nevertheless covers a broad swath of our intended goal. 

%We identified a set of four representative scenarios for the non-place traveller: the daily commuter, the long-distance traveller, the journaling traveller, and the quantified-traveller. We then explored opportunities for social computing for the non-place traveller using these four scenarios. The result of our efforts is the design of Journeys \& Notes, a location-based social mobile platform designed with non-place travellers in mind. 

%We deployed this app in two stages: first we recruited participants to pilot the app for two weeks, then we released the app to the public, gathering more than 18,000 users who logged about 10,000 journeys and wrote 2,000 notes over the course of seven months. 

The main contributions of this work are:

\begin{itemize}
\item The design and implementation of Journeys \& Notes, an Android app that lets people check-in to their journeys.
\item Two 1-week-long user studies (6 people, and 15 people) of how people adopted the system, and how their use of the system impacted their perceptions of travel and non-places.
\item A large-scale field study on 9,435 participants providing insight into usage patterns, and the types of messages people write on their journeys.
\end{itemize}

%Finally, we hope this work can serve as a case study for how system design and analysis in HCI can be motivated by and guided by social theory.

\section{Background and Related Work}

%We describe the background and prior work that motivated Journeys \& Notes, and we also contrast it with existing commercial applications.

%\subsection{Location Based Social Application}
Check-in apps emerged around 2003 to enable a virtual social experience anchored around physical places \cite{Lindqvist:2011}. These apps allow users to broadcast their presence at venues, that is, the places they go to. Venues include restaurants, bars, offices, apartment buildings, homes, museums, parks, movie theatres, shops, and caf\'{e}s. Dodgeball, GoWalla, Foursquare and now Swarm are some of the apps that embodied this concept of checking-in to a venue. More recently, Facebook has adopted this as a feature that people can use to attach a location to their posts.

%\subsection{Spaces: Places and Non-Places}
Designers of social computing technologies have grappled with the concept of places for quite some time. In 1996, Harrison and Dourish \cite{Harrison:1996} argued for the importance in distinguishing between ``places" and ``spaces," with a special emphasis on the virtual. Ten years later, Dourish \cite{Dourish:2006} restates these concepts in the context of spatial technologies.

Places often create communities around them \cite{putnam_bowling_1995, oldenburg_great_1989}. Place check-in apps have begun to leverage and augment these venue-based communities \cite{mccarthy2009supporting, schwartz2014online}. Conversely, non-places are not typically conceptualized as vibrant spaces where communities take root. However, we have observed instances where online communities anchored at non-places have shown signs of emergence. 

%Rem Koolhaas in his article ``Junkspace" explored how mass produced architectural forms and building materials are producers of the same sense of placelessness that Aug\'{e} describes \cite{koolhaas2006junkspace}. Although our work is not explicitly concerned with architectural elements, the form of the spaces the traveller moves through likely impact the notes that they leave.

Previous work has explored the design of technologies to connect people with one another while riding public transportation. For instance, researchers developed Trainroulette, an app to promote ``situated in-train social interaction between passengers'' \cite{camacho2013trainroulette}. The researchers found that people were interested in knowing who shares the rides with them, but wanted to do it semi-anonymously (exposing only certain aspects of their identity). Similarly, Belloni and colleagues \cite{belloni2009see} explored the use of mobile technologies in ``transitional spaces.'' Their work focused on the design of a location-based friend finder that displays any of the user's friends that are in the same subway train. The researchers found that users wanted the ability to ``invisibly'' log in to the system. This need for identity opaqueness inspired the design of our app's identity system. 

Other work has looked at designing applications to increase social engagement with the physical world. Rosner et al. designed an application that let people convert free form doodles to sharable walking routes on maps \cite{rosner2015walking}. Cranshaw et al. designed an online community of people documenting their experiences at places through sharable city guides \cite{cranshaw2014curated}. Overall, a meta-analysis of pervasive technology and public transport \cite{camacho2012pervasive} proposed the development of applications that not facilitate only more efficient journeys, but also more enjoyable ones that people look forward to. This is what we set out to do.

%\textbf{ADD PAPERS WE NEED TO CITE HERE: **************} 
% CITE SOME PAPERS THAT ROSNER2015 cites

\nocite{kim2015travel, kraut2012building, preece2000online}

\section{Scenario-Based Design for Non-Places}

Our investigation of non-places as an unexplored domain in social computing began with an informal survey of how existing social computing systems and apps are being used or re-purposed for use in non-places. In surveying this landscape, we considered all-purpose social networks such as Facebook, check-in apps like Foursquare, fitness apps like Strava, tracking apps such as Moves, travel apps such as TripAdvisor, and commuting apps such as Waze. As we considered each technology, we thought about how the features and usage scenarios of these existing services might be adapted by the non-place traveller, perhaps to help foster a disappearing sense of community, or to augment a transient space with virtual experiences of place. 

Inspired by formal design methodologies such scenario-based design \cite{carroll2000making, CarrollFiveReasons, Nielsen:2002, rosson2009scenario} and persona-based design \cite{pruitt2003personas, Chang:2008, blomquist2002personas}, we synthesized our findings into a set of four scenarios that we felt represent broad yet diverse archetypes of the non-place traveller, while speaking to the opportunities, challenges, and implications of life in non-places. Here we highlight how these scenarios helped inform our design process, offering a lens through which we can empathize with the lives, emotions, and social computing needs of a non-place traveller. 

% Here we tell the stories of \emph{Joan}, a bus commuter who finds solace in the familiar faces and places along her daily route; \emph{Aaron}, a world-traveller who spends more time passing through airports than he does at home, \emph{Jacque}, a data-obsessed, quantified-self enthusiast who counts, tracks, records, and analyzes every detail of every trip he ever takes; and \emph{or}, an observant documentarian of the prosaic but beautiful wold around her, whose notebook never leaves her side as she explores her neighborhood. 

%Here we consider scenarios for the everyday commuter, the frequent flyer, the quantified traveller, and the journaling traveller. 

\subsection{The Everyday Commuter}
% Waze, repurposing 4sq

{\sl
Whether they travel by car, by bus, by train, by bike, or by foot, the one thing most everyday commuters have in common is the regularity of their trips. Especially in larger urban cores, this often translates to either time spent in the car stuck in traffic, or sharing the same bus or train with many of the same familiar strangers they see each day.
}

Recently, apps such as Waze \cite{Silva:2013} have successfully harnessed the excess capacity of people's free time during peak commuting hours by using people's smartphones to crowd-sourcing traffic conditions. 
Although its goal is utilitarian, Waze built an online mobile social community around commuting drivers stuck together in traffic, all just trying to find a way around. This shows how social computing can be designed with the needs of the everyday commuter in mind.

%\subsubsection{Example Persona: Joan}
%Recent decisions by the county to decrease bus service to the northernmost neighborhoods of the city concerned Joan. For the last 4 years, she has taken her faithful bus 32X from her starter home to the city center. Each day, like clockwork, she climbs the three steps onto her bus, greets her driver, and sits next to the window. Joan was among the first riders to get on the bus and after four years, she has developed a pseudo-anonymous relationship with her fellow passengers. Mr. Burgundy, named after his classic Merlot ensemble, gets on at stop 4 near the park; Mrs. Nest, so named for the bun of hair piled on her head, joins the adventure outside of the library with her latest pile of books; Henry and Henrietta, two school-aged children, are scooped up near the grocery store and dropped ten stops later at the neighborhood school. Joan knows them all, maybe not by name, but she knows where to expect them. They are a part of her life, familiar strangers. She begins to wonder about the small, transient community she has developed: what will happen when services get cut? Sure, people will find new routes, perhaps walk farther to access public transit, but who was really looking out for the faithful riders of 32X?

\subsection{The Frequent Flyer}
% Long distance Checking in airports on 4sq. Airport Tango.
{\sl
Red-eye to London. Dinner at an airport bar in Istanbul. Just enough time for a coffee before the early flight to Fez. The frequent flyer knows airports. There are no familiar strangers in the airport, only familiar types -- the business traveller, the retiree, the honeymooning couple, the college student returning home, the jet-setting socialite -- the frequent flyer can spot them all with barely a glance.
}

\begin{figure}
\centering
  \includegraphics[width=0.98\columnwidth]{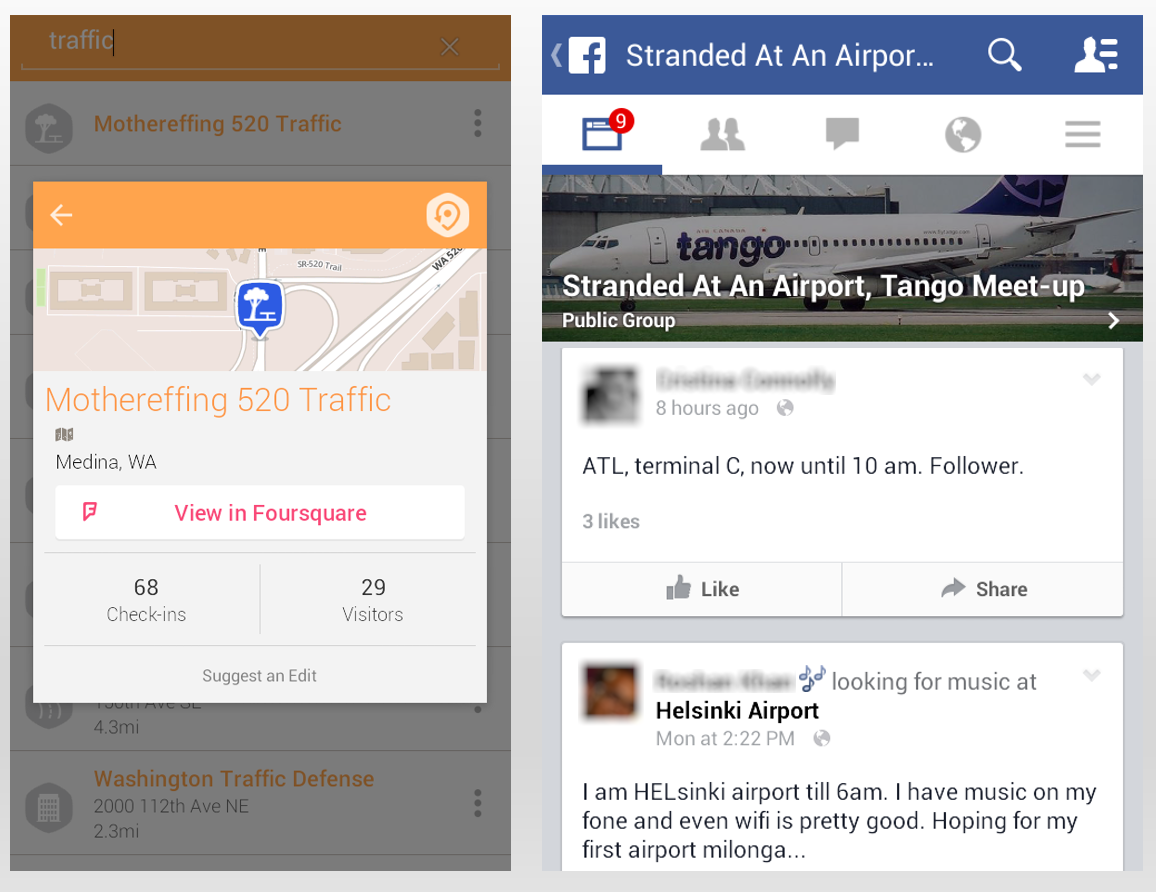}
  \caption{(Left) Swarm  venue for a recurring traffic jam in a highway.(Right) Facebook group for meeting tango dancers at airports.}~\label{fig:4sqtraffic-tangofb}
\end{figure}

Airports are among the most highly checked-in-to venue types in Foursquare and Facebook. Sometimes, flyers check-in on departure to let friends know they'll be out of town. Others check-in to boast about the interesting places they're travelling to. Checking-in upon arrival in an unfamiliar place offers the opportunity for serendipitously bumping into nearby social connections. A Facebook group called Airport Tango\footnote{\url{https://www.facebook.com/groups/348726108583892/}} has over 8,000 members. Whenever a group member has some layover time in an airport, they post to the group looking to connect with other members so they can dance the Tango while they wait for their flights (see Figure~\ref{fig:4sqtraffic-tangofb} Right). Similarly, Foursquare users often create fake venues to check in while stuck in traffic jams (see Figure~\ref{fig:4sqtraffic-tangofb} Left).  Check-in apps and other forms of social computing are regularly being used by travellers to make airports a little more social. 

%\subsubsection{Example Persona: Aaron}

%Aaron's familiarity with red-eye flights has grown over the years as he transformed into the go-to guy for international business travel. No family, a small rental on the Lower East Side, and very little waiting for him at home other than his air plant. Home, a concept that has begun shifting over the last few years, seems to be a transient place. Aaron leaves a piece of himself behind in each city he visits, whether it's a new locale for this frequent flyer or a semi-regular haunt he's visited before. Shared meals with locals, evenings watching the spice market come alive, walking through the overgrown hedges of a garden long forgotten, Aaron finds a home in his adventures and a community in the people he spends his time with abroad. His familiar faces are thousands of miles away, waiting for his next big trip, and eager to share in another adventure. 

\subsection{The Quantified Traveller}
% Moves, Strava, Loom 
{\sl
Counting -- a cousin to logging, tracking, following, and recording -- requires extreme consideration and attention to detail. The quantified traveller might count calories, steps, rides, smiles, records, sodas, hellos, sighs, words spoken, liters drunk, grams consumed, emails sent, texts sent, etc. -- you know, the things that prove useful when compiling the annual report.
}

The quantified-self movement seeks to empower individuals with increased capabilities for tracking and expressing their personal activities. Some of the initial commercial successes in this space have been glorified step counters, allowing people to estimate how much they walk in a given day. Recent quantified-self applications are starting analyze the trips people take. For example, Strava is a smartphone app for competitive bicyclists and runners that tracks their routes and times \cite{Cintia:2013}. It goes well beyond simple GPS path tracking, however, implementing a leaderboard that shows the best recorded times on all road segments. Moves is another quantified-self smartphone application that builds a beautiful visual journal of the users' activities by automatically identify the venues that they stop at and the modes of transit used to take them between venues. 

%\subsubsection{Example Persona: Jacque}
%The extreme consideration and attention to detail required for careful counting are qualities at which Jacque excels. Each day's records of his personal data events -- calories, steps, rides, smiles, sodas, hellos, sighs, words spoken, liters drunk, grams consumed, emails sent, texts sent, etc. -- proves useful when he compiles his annual report. The report, if unraveled, contains a detailed account of how to live the life of a 31.23 year-old male in Atlanta. While the details may not leave many on the edge of their seat, Jacque feels a certain success in pioneering faster walking paths that maximize his elevation gain and give him obscure anecdotes about having once caught the elusive, and limited, midnight 10 bus to Midtown.

\subsection{The Journaling Traveller}
% Journaling Apps, Cloud based notebooks, Day1 (sensing), repurposing of apps (4sq, Yelp) for journaling 

{\sl
The journaling traveller documents the details they feel are often overlooked, carrying their stories with them -- on fronts of notebooks, side pockets, marking events on calendars. Documenting the world as they see it drives the Journaling traveller to write at length about how their neighborhood is changing, about the gentleman who walks the park every morning at 7:19am, and about the most recent paint change on the sushi restaurant next door.  
}

Documenting journeys is an age-old tradition. One of the earliest travelogues recorded is the one by the Greek geographer Pausanias from second century A.D. Today, the explosion of social media applications has motivated people to document every aspect of their lives, and journeys are not an exception.  People document them with photos on Instagram, or by writing long and detailed notes on Yelp or TripAdvisor about their journeys. Journaling-specific mobile applications like Day One have millions of people using them diligently every day. All of this shows the desire both for journaling travels and for mobile apps to do so.    

%\subsubsection{Example Persona: Eleanor}
%Casual observances in the structures lab transform into small stories jotted down on sticky notes. The aptly titled ``An Ode to the Settling of Cement" catches the eye of anyone willing to give the electric pink note a prolonged glance. Eleanor carries her stories with her -- on fronts of notebooks, side pockets, marking events on her calendar -- in order to document the details she feels might be overlooked. Her passion to write grows each day. Documenting the world as she sees it drives Eleanor to write at length about how her neighborhood is changing, about the gentleman who walks the park every morning at 7:19am, and about the most recent paint change on the sushi restaurant next door. It is through her fresh eyes focused on the ordinary that she finds others noticing the same thing, a community of people capturing the lesser told stories of the everyday. 

\section{Journeys \& Notes System Design}

In this section we describe the key elements of the Journeys \& Notes system design. We link specific designs and components to the various facets of social life in non-places that motivated them, often using our four scenarios as a connecting point of reference.

\begin{figure}
\centering
  \includegraphics[width=0.49\columnwidth]{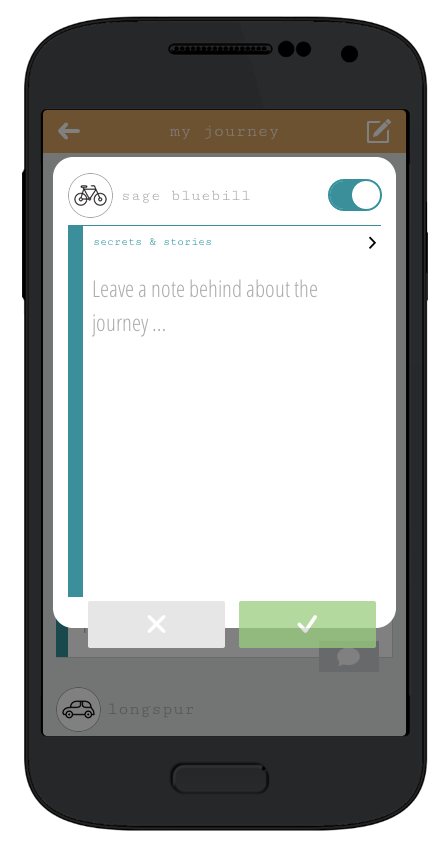}
  \includegraphics[width=0.49\columnwidth]{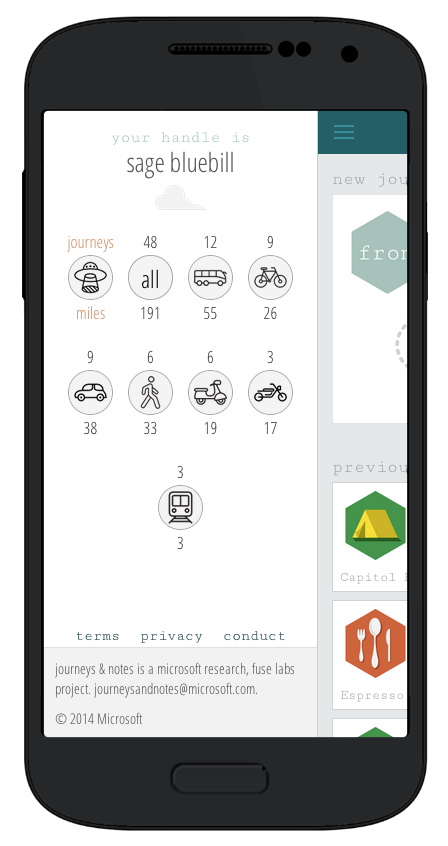}
  \caption{(Left) The \emph{Notes} view. Users can leave a note behind on their journey for others to see. (Right) The \emph{Stats} view. Users can see summary stats of all the journeys they have taken, grouped by mode of transit.}~\label{fig:notes}
\end{figure}

\subsection{Overview}  
  % previous journeys
  % new journey
  % current journey
  % mode of travel
  
%The starting point of our user experience design is the location check-in. 
Journeys \& Notes extends the concept of a check-in \cite{Lindqvist:2011, cramer2011performing} beyond fixed-point venues, allowing users to check-in to the \emph{journeys} they take. By specifying the endpoints of their journey and their mode of travel, users unlock a virtual social experience where they interact with other travellers also on the journey, discover notes that others travellers have left along the way, and they can leave behind their own notes for others to find. 

The app is comprised of two main screens: the \emph{Home} screen, and the \emph{Journey} screen. From the home screen (see Figure~\ref{fig:mainscreens} Left), there are four possible actions that the user might take: (1) they can check-in to a new journey; (2) they can review their previous journeys, possibly checking-in to one of them; (3) they can return to see details of their current journey; and (4) they can review statistics about all their journeys by mode of transit (see Figure~\ref{fig:mainscreens} Right). 

When the users tap the ``new journey" area, they are prompted to specify the origin, destination, and mode of transit to complete their journey check-in. When checking-in to a previous journey, the user is only required to specify the mode of transit (see Figure~\ref{fig:modes} Left). Previous journeys are presented on the Home screen as a visual list of cards for each journey, depicting the origin, destination, and summary statistics about past mode-of-transit choices on that journey. If a user is currently checked-in to a journey, they will see ``current journey" section on the home screen, with a card depicting the details of the journey they're checked into. In all cases, these actions take the users to the \emph{Journey} screen for their current journey.

\begin{figure}
\centering
  \includegraphics[width=\columnwidth]{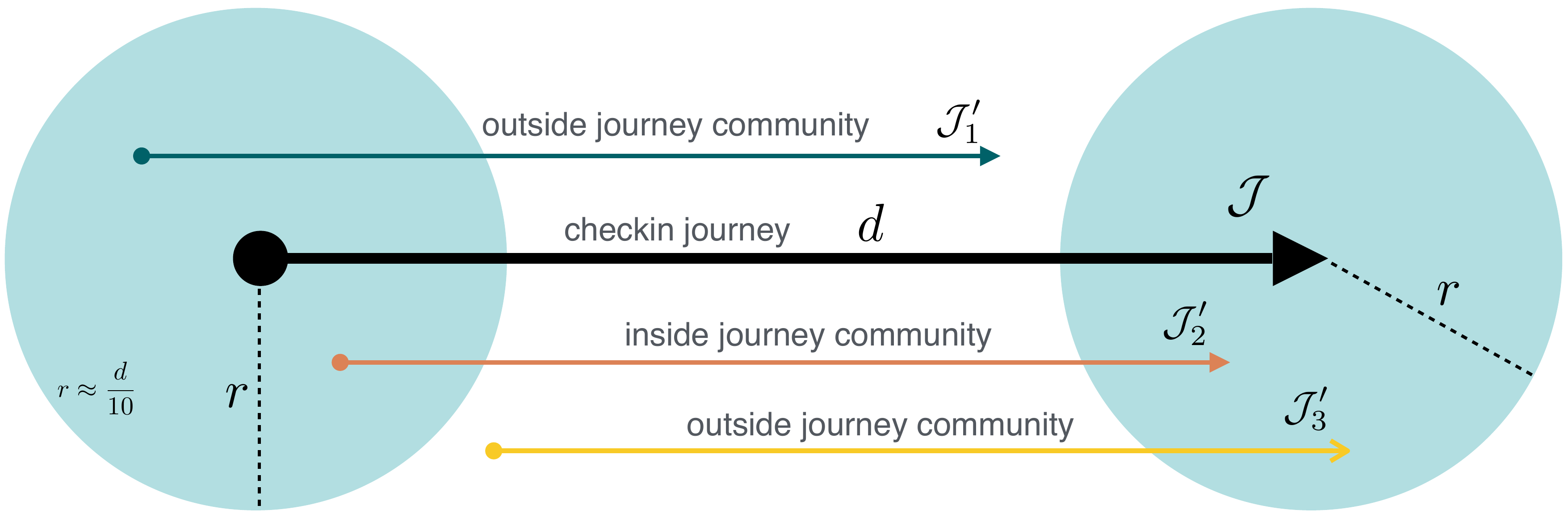}
  \caption{Nearby journeys are bundled together to create a community. $\mathcal{J}$ is the check-in journey of distance $d$, which defines a community radius of $r \approx \frac{d}{10}$. Only notes $\mathcal{J}'_2$ are visible to $\mathcal{J}$, because$\mathcal{J}'_2$ starts within $r$ of $\mathcal{J}$ and ends within $r$ of $\mathcal{J}$.}  ~\label{fig:journey_community}
\end{figure}

\subsection{Journeys}

A \emph{journey} is fully specified by its \emph{endpoints} -- the origin and  destination of the user's trip. In the system, endpoints are literally points on the map (latitude and longitude coordinates), so that two users will be on identical journeys if they start at the same location and finish at the same location, independent of the paths they take while en-route. The user provides their journey endpoints in a text field by specifying either a \emph{venue} (e.g. a named points of interest, like ``The Statue of Liberty") or street addresses. The system geolocates the user's input, resolving the endpoints to a coordinate by referencing the Foursquare API (for venues) and the Bing API (for addresses). To enable ease of free text input on a mobile device, as the user types their endpoints  the system generates ``typeahead" auto-suggestions of nearby venues and addresses based on the current device location \cite{aniszczyk_twitter_2013}. %\footnote{\url{http://twitter.github.io/typeahead.js/}}

  % bing and foursquare
  % location context sensitive
  % autocomplete for venue / address
  % endpoints are hidden to others

\subsection{Journey Check-ins}

The central experience of Journeys \& Notes is the \emph{journey check-in}, the process by which users expresses their presence on a particular journey. To check-in to a journey, a user must give their origin and destination (determining which journey they are on), and they must choose their mode of travel from a menu of travel icons (See Figure~\ref{fig:modes} Left). 

When a user checks-in to a journey, they are taken to the Journey screen, where they are greeted by a randomized ``welcome" message that we crafted to reinforce the user check-in behavior, and adding variety, playfulness, and gamification aspects to the user experience. We designed three types of welcome messages. The first type shows high level statistics about the journey, including any notable milestones, how many times the user has travelled the journey, and how many other people have travelled on the journey too. The second message type presents a travel-related haiku for the user's enjoyment. The third type shows ``fun facts" about the journey (for example how long would it take a bird to travel the whole distance). In later sections, we will delve into the design and motivation for showing these messages.  Figure~\ref{fig:compose} (Right) shows one example of the welcome screen.

\subsubsection{Journey Community}

The Journey page is a forum that asynchronously connects users on similar journeys, allowing people to leave behind notes for their fellow travellers to read, and to read and respond to the notes left behind by others. These notes are only visible when a user is on that journey, creating an explicit bond between the virtual content of the journey, and the physical journey itself. 

This design augments the space of the traveller with a persistent community, complete with affordances to interact with this community outside of the physical realities of the non-place. This design is one of the most direct vehicles by which we hope to impact the experience of non-places.

Each check-in journey defines its own forum members based on the journey endpoints. Roughly speaking, all other journeys that start \emph{near} the check-in journey's origin, and end \emph{near} its destination are bundled together to define the journey's forum. We scale this bundling so that the longer a distance covered by the check-in journey (as the crow flies), the wider the community of bundled journeys. Specifically, if the check-in journey $\mathcal{J} = (a,b)$ travels from point $a$ to point $b$ and is distance $d$, we define the community radius $r = (d / c)$ to be a constant fraction of the distance for some $c > 1$. As $d$ varies, we further restrict $r$ so that it has fixed minimum and maximum values $\alpha$ and $\beta$. Then the community of journeys bundled with check-in journey $\mathcal{J}$ is $C(\mathcal{J}) = \{ \mathcal{J}'=(a',b') : d(a,a') \leq r \text{ and } d(b,b') \leq r \}$. In our implementation we set $c$ to 10, $\alpha$ to 100 yards, and $\beta$ to 30 miles. That is, $\mathcal{J}$ is bundled with other journeys that start within a radius equal to 1/10th of its distance and end within 1/10th of its distance (with an additional minimum radius constraint of 100 yards and maximum radius of 30 miles). This is outlined with an example in Figure~\ref{fig:journey_community}.

Having a community radius that scales with the length of the journey allows Journeys \& Notes to be equally useful for the long distance business trips, where you would want to encompass the entire cites or each end, and for the many intra-city adventures of the everyday commuter, where radius should capture just a block or a neighborhood. 
%Travelling over longer distances, say from New York to San Francisco, one would want to be grouped with other travellers that start anywhere in the New York region, and end anywhere in the San Francisco region. Similarly, if one is travelling between two neighborhoods within the same city, one would want a much more localized community. All of this happens invisibly in the user experience: when they check-in to a journey, they simply see their fellow travellers on similar journeys.

  % journey community
  % journey community radius
  % endpoints are hidden to others
  % 
  
%  In doing so
%- attempting to nurture community structure in spaces where community is latent
%- giving people tools to connect in spaces where it may be difficult or impossible to otherwise

%\begin{figure}[t]
%\centering
%\includegraphics[width=\columnwidth]{figures/checkins_by_mode.pdf}
%\caption{The number of journey check-ins observed during the field deployment across different modes of transit. %Cars (34\%), Airplanes (17\%), Trains (14\%), Busses (12\%), and Pedestrians (7\%) comprise the majority of %check-ins.}
%\label{fig:checkins_by_mode}
%\end{figure}

\subsubsection{Mode of Travel}

\begin{figure}[t!]
\centering
  \includegraphics[width=0.41\columnwidth]{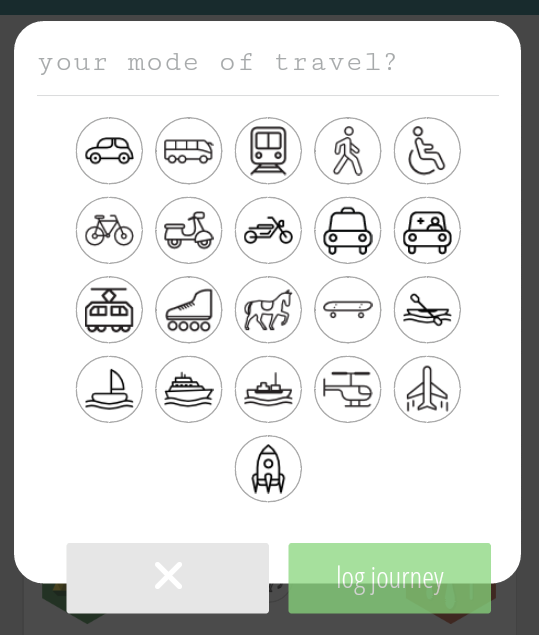}
  \includegraphics[width=0.57\columnwidth]{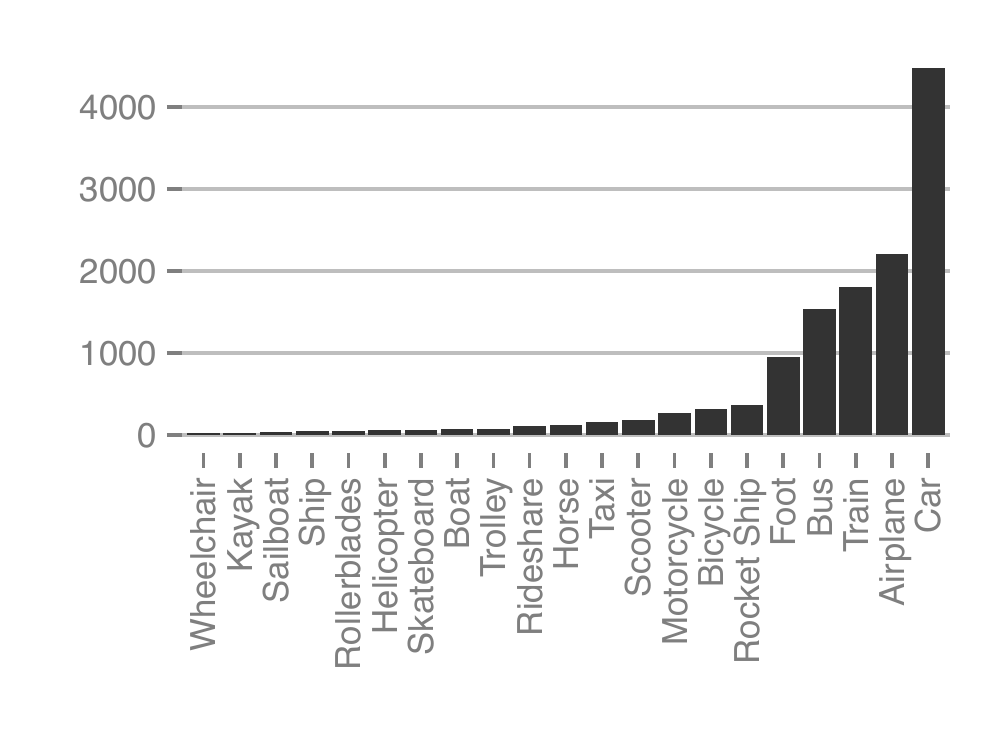}
  \caption{(Left) Supported modes of travel. (Right) The number of journey check-ins by transit mode observed during the field deployment. Cars (34\%), Airplanes (17\%), Trains (14\%), Buses (12\%), and Pedestrians (7\%) comprise the majority of check-ins.}
  ~\label{fig:modes}
\end{figure}

Asking users to select the mode of travel in order to check-in to a journey serves three very different purposes. First, by forcing travellers to think of their mode of transportation as not just a means to the end of their journey, but as a critical part of their identity in the online community (their avatar), we are turning role of the mode of travel on its head, from something that encourages individuality and disorientation in the physical world, to symbol that breeds sociality and presence in the online community.

Second, their mode of travel provides an important piece of contextual information that can be used to help foster online community growth. To surface this information, we use the mode of travel icon as the user's avatar in the journey forum. This increases travellers' awareness of other possible modes of travel to get where they are going; for example, drivers become more aware of the cycling community that exists along their commute. It also encourages communities to form within a given mode of travel; for example, bus riders can quickly scan posts from other bus avatars and identify with other bus riders' travelling experience.

Finally, data about how people get to where they are going is extremely valuable civic data that could be used to better inform city transit planning. Collecting these data at a large scale is very difficult for cities and organizations. Developing a scalable approach for approximating or modelling transportation patterns within a city could lead to significant civic innovations.

  % reinforce community features
  % bring awareness of different lifestyles
  % help with information filtering
  % frame the context of the discussion

\subsection{Notes}
  % note sections
  % comments
  % feed
  
\begin{figure}[t!]
\centering
  \includegraphics[width=0.49\columnwidth]{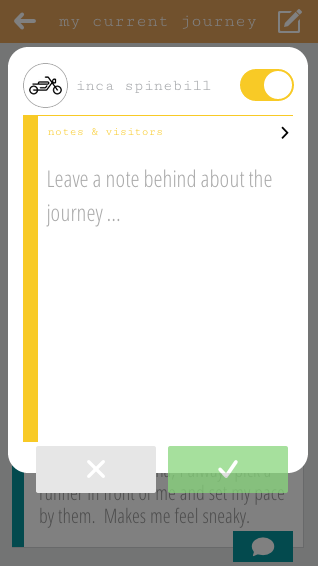}
  \includegraphics[width=0.49\columnwidth]{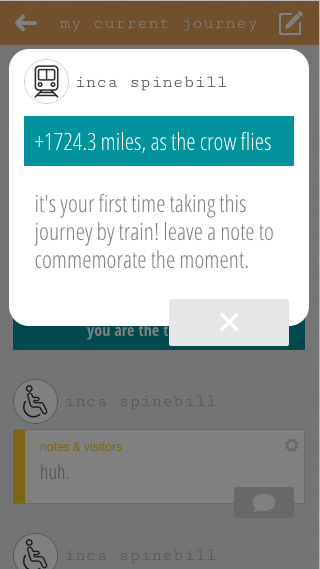}
  \caption{(Left) The note \emph{compose} window. Tapping the toggle in the top right switches between anonymity and pseudonymity. Tapping the sub-header will step through the five different note sections. (Right) A \emph{welcome} pop-up appears immediately after a user checks-in to a journey.}~\label{fig:compose}
\end{figure}
  
Leaving notes on a journey is relatively straightforward. Simply tapping a "compose" button opens up a note builder popover window, which guides users to constructing their notes. Notes are plain text and are restricted to 250 characters. The compose view is shown in Figure~\ref{fig:compose}. The author has the option to categorize the note with one of five categories: \emph{Notes \& Visitors}, \emph{Secrets \& Stories}, \emph{Love \& Hate}, \emph{Missed Connections}, and \emph{Tips \& Tricks}. By scaffolding the compose window with note topic sections, our goal is to gently focus the discussion, so that people are not turned away by the cold start problem of a blank page with no inspiration for what to write.

Notes appear in the Journey page in a feed displayed in reverse chronological order. Different categories receive different color treatments, allowing for visual information filtering when browsing the feed. When a note in the feed is tapped, it opens up a detail view for that note, which displays any comments that have been left, and exposes an interface for composing a comment on the selected note.

Note are not only the primary means by which people interact with their fellow travellers (by reading the notes of others, and leaving comments behind), but simply by virtue of being asked to write about it, notes encourage the traveller to be mindful of the journey, and create a sense of history and culture from the previous travellers.
  
\subsection{Pseudonymity and Anonymity}
  % familliar strangers
  % anonymity in cities
  % bird names

\begin{figure}[t!]
\centering
  \includegraphics[width=0.49\columnwidth]{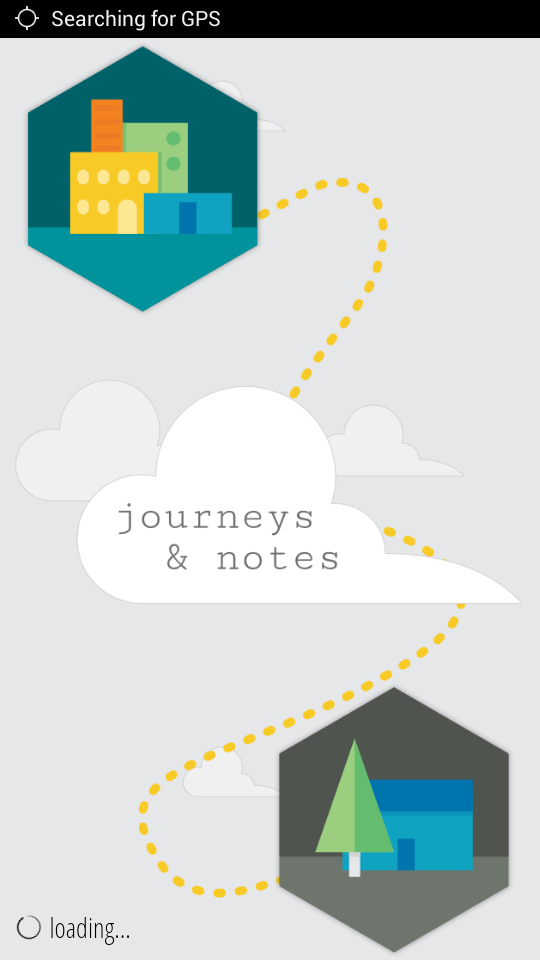}
  \includegraphics[width=0.49\columnwidth]{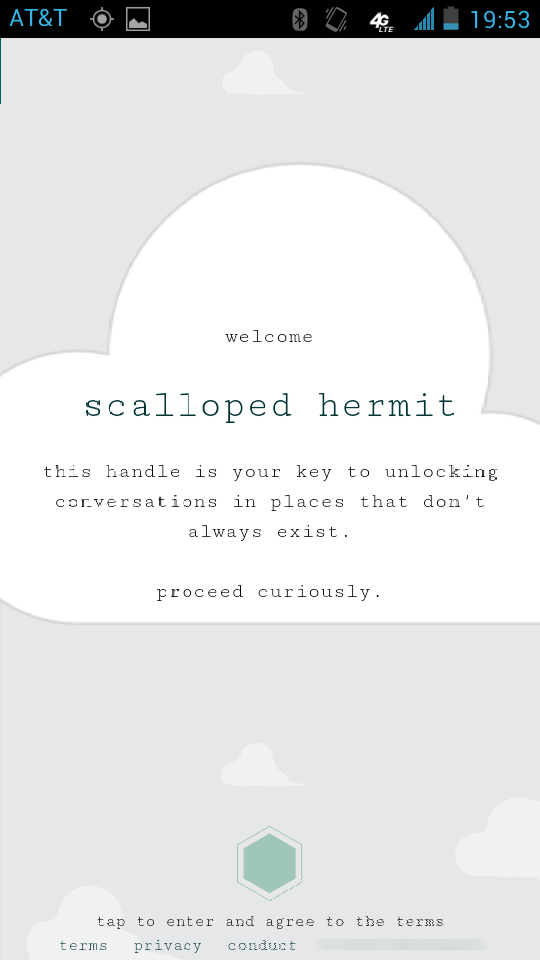}
  \caption{(Left) Journeys \& Notes loading screen. (Right) A random pseudonym given to a first time user. }~\label{fig:loading}
\end{figure}

There are three defining aspects of our design around user profiles and identity: (1) users are given random system-assigned usernames, or \emph{pseudonyms}, the first time they use Journeys \& Notes; (2) pseudonyms are persistent, they cannot be changed, and they are visible on all the notes they leave behind; and (3) when users are composing a note, they can decide (with the switch of a radio button) to make the note completely anonymous, thereby not associating that note with their pseudonym. 

Our pseudonyms are derived from  naming schemes developed for bird species common names. Although there are only about 10,000 species of birds \cite{list_2015}, we developed a method for creating a suitably large name-space of pseudonyms that \emph{sound like} bird names, but may not be, by expanding on typical structure used in naming strategies. Each species common name must feature the common name of the family (e.g. sparrow, wren, warbler). However, the family name is typically modified by (1) attributes describing the physical characteristics (e.g. yellow-rumped cacique); (2) geographic terms denoting where the bird can be found (e.g. the Eurasian nuthatch); or (3) landscapes or habitats where the bird roosts (e.g. mountain pygmy owl). By taking the Cartesian product of the constituent sets of descriptors in each of these three cases, we created a name space of well over a million possibly fake but convincing bird common names (e.g. scarlet crested wren). We felt having bird names as pseudonyms was a whimsical and playful design that might appeal to users like the journaling traveller, who sees beauty in the simple things around her. Figure~\ref{fig:loading} illustrates the new user on-boarding screens, highlighting the pseudonym assignment. 

Much of our thinking about identity and anonymity was influenced by the needs and experiences of the city-dwelling public transit enthusiast, such as the everyday commuter scenario. Many who live in large cities have high regard for the anonymity that comes with city life, a somewhat paradoxical byproduct of sharing a small physical area with a great many people \cite{jacobs:1961}. They also value the structure and regularity of the daily commute -- in particular the familiar strangers, those faces in the crowd that they recognize and who may even occupy a significant portion of their lives, yet whom they have never had any deep interactions with. 

Stanley Milgram was one of the first to write about the role of familiar strangers in urban milieu. He describes the phenomenon as a distinct category of relationship, providing support and value, and also requiring maintenance in the same way that family, friends, or colleagues do \cite{milgram:1977}. Like Paulos and Goodman's Jabberwocky \cite{paulos:2004}, our design of Journeys \& Notes does not attempt to alter existing familiar stranger norms or relationships. Rather, we hope to accentuate familiar strangers by building technology that enables new interactions and experiences that highlight their unique role in urban life. 

Specifically, our decision to use assigned pseudonyms was intended to preserve urban anonymity. For the everyday commuters who regularly share close quarters with many of the same fellow travellers every day on their bus, train, or shuttle commutes, we believe it is important to maintain the same feelings of anonymity in the application as is experienced in the physical world. Our use of persistent random pseudonyms, in contrast to pseudonyms that are refreshed with each new post \cite{lee_shh_2014},
%\footnote{Recent popular anonymous geolocated chat apps such as Secret, Whisper, and YikYak use random usernames that refresh with each new post.} 
allows users to maintain familiar stranger relationships in the application by noticing and following the posts of familiar pseudonyms that they might recognize from journeys past.

\subsection{Quantified-Self and Gamification}
%trailbrazer
Just as place based check-in applications use self-quantification and gamification as motivating forces that drive usage \cite{Lindqvist:2011, zeynep2011gamification, cramer2011performing}, we believe there are similar opportunities for non-place based check-ins. First,  inspired by a similar feature Foursquare used to have, we give people the title of ``trailblazer'' when they are the first to check-in to a journey that no one else has checked-in to before. Our goal is to encourage people to explore new journeys for themselves and for the system as a whole. Furthermore, inspired by the examples we described in our scenario like Waze and Strava, we decided to incorporate journey-based statistics about the user. We surface these statistics to the user in a summary panel on the Journey page. We first maintain simple counts for the number of times the user has taken each trip. Additionally, we maintain counts for the number of times the user has checked-in with each mode of transit, both in total and for each journey. We also measure cumulative miles traveled (as the crow flies) for each mode of transit, again both in total and for each journey. The total cumulative mode of transit stats is displayed on the user profile page, see Figure~\ref{fig:notes} (Right).

  % number of times taking each journey
  % number of times taking each mode in a journey/total
  % number of miles taking each model in a journey/total
  % trailblazer
  % journey stats
  % journey trivia
  % first time for each mode

\subsection{Design Aesthetics and Poetic Elements}
  % haikus
  % visual design language
  % message in a bottle experience
  % poetry on busses
  
The idea of leaving a digital note behind on a journey evokes feelings of hiding a message in a bottle and tossing it out to sea --- the action itself is poetic --- writing a message not knowing who, if anyone, will ever see it. From the visual design language of the application, to the use of bird names as pseudonyms, to the natural imagery scattered throughout, we've tried to imbue much of the poetics of this moment into the look, feel, and voice of the application. These poetic elements encourage travellers to be mindful and present in their surroundings, perhaps nudging them to notice aspects about the journey that they otherwise might have missed. 

The most overt invocation of poetics in our design was our use of haikus to welcome users to a journey and priming them, to think about the type of content they might leave behind as notes. We hired three creative writers on oDesk\footnote{\url{http://odesk.com}} to produce haikus exploring the core concept of the app: the journey through non-places. Each writer composed 15 haikus, three haikus inspired by each of the five note sections. 

Here is one inspired by \emph{Missed Connections}:
\begin{quote}
Motor hum, grease smell \\
and there you are again, but \\
long, the road between
\end{quote}

%The usage scenario about a journaling traveller was one of the main inspirations for including these whimsical elements in the design. Seeing journey-related haikus might inspire such travellers to write some poetry of their own. Perhaps they'll even leave some poems behind as notes for others to find.

\subsubsection{Seeded Content}  
%%% HERE IS THE WRITERS WE HIRED TO SEED CONTENT
We also hired a number of writers who were familiar with a number of major metropolitan areas of the United States to seed the app with pre-existing 896 notes as one way to bootstrap engagement. We asked the writers to write notes for common journeys of people in those metro areas.  
%We saw the app as a conduit to deliver a user experience composed of both functionality and content. 
Below is an example of a note written by one of these writers:

%\begin{itemize}
 %\item ``We were riding the 49. I had just gotten off work and was wearing a stuffy button-up and tie, while you were wearing jeans and a tank. Your legs brushed against my knees when you got off at Broadway and John."
% \item 
\begin{quote}
``Every day during my lunch break I take a run from the bottom of Boston street up the hill and I draw a chalk line to see where I am after 10 minutes. I'm inching my way up, guys, anyone wanna race?"
\end{quote}
 %\item ``My ex and I used to come here all the time. Still one of my favorite places, but more than a little bittersweet now. Wonder if this feeling ever fully fades."
%\end{itemize}

 %``Take a Car2Go or Uber instead of a cab or bus. Avoid the crazies on the bus for only a few bucks more."
 %``Find a way to live close enough to where you work, or find a way to make your commute time count. Took me years to figure this out; so much more productive now."

\subsection{Implementation}

Journeys \& Notes is implemented as a client (Android application) and server (REST API), where the server's role is exclusively to persist the actions of the user to a remote datastore, facilitating sharing and interaction across multiple distributed clients. The datastore is implemented with a spatially enabled PostGIS/PostgreSQL database, allowing for efficient execution of queries with spatial joins. Our REST API is implemented in Python with the Flask framework. We will not go into any further details about the system architecture.

\section{User Studies}

We conducted two rounds of small scale user studies of individuals who used Journeys \& Notes for one week with a total of 21 participants. 
In study 1, we recruited 6 participants from the Seattle area through TaskRabbit\footnote{\url{http://taskrabbit.com}}. 
%\cite{Teodoro:2014}.
We conducted an hour long, in-person pre-study interview with each participant, to get a sense of their normal everyday transportation patterns, how they got around the city, and what they typically perceive as they travel. After this discussion we gave them an overview of how to use Journeys \& Notes. They were asked to use Journeys \& Notes for one week, to check-in on each journey they take, and to leave 3 notes behind on their journeys per day. After this week, we conducted another in-person, 1 hour long post-study interview.
In study 2, we recruited 15 people from the Seattle area through Craigslist\footnote{\url{http://craigslist.org}} to use Journeys \& Notes for one week with the same usage requirements of study 1 participants. In this round, however, our evaluation was through two online surveys (pre- and post-study), featuring several free response questions about their experience.

Our goal was two-fold. First, we wanted to better understand the landscape of how people conceptualized the journeys they take and people they encounter along the way. Second, we wanted them to use Journeys \& Notes for a week, so we could get a high level understanding of which aspects of the system people seemed to respond most to, and which were features were not as robust. The version of the app that we tested in these trials was more primitive than the one described above. It did not have as many quantified-self features that counted trips and mileage. We transcribed the interviews and identified common themes and important insights that emerged, which we synthesize in the discussion below.

\subsection{Anonymity and Other People}

Because the content was sparse and geographically siloed across the city, participants had limited contact with others in the study, however there was some. One participant remarked laughing at posts he found from other people: ``Some of the missed connections are really funny. There was one guy that lost his blue velvet jacket somewhere up there [laughter]... he was looking for it. He lost it on a bet. That was hilarious!" Although most people did not post anything fully anonymously, some of those who came into contact with others' posts seemed more thoughtful about using fully anonymous postings, often when saying things that they didn't want attributed to their pseudonym. One participant posted all of her content anonymously. She was a regular bus rider, and was concerned about people identifying patterns in her posts and being able to guess her real identity, so she decided to be entirely anonymous. 

%\item[Community, Anonymity, and Familiar Strangers]
%"Again the same thing when I was working down in SoDo. You would recognize the same people, because my timeline was the same, and even when it wasn’t I found that interesting. There was some people who like me, sometimes didn’t make the 15 minute earlier or 15 minute later, but you would just being to recognize these folks. I suspect people who do this everyday for a year, two years, three years, four years, but they still don’t interact that much. even if they start and end their destination in the same place, that’s the thing that I thought was interesting. [bangs hand on the table]"

%The further away from downtown, it tends to be more racially diverse, more women, except for the airport commuters, who are varied because this is a fairly racially diverse town. But the further south you go as you’re headed towards tukwilla and all that stuff, it just becomes more racially diverse and farm more women, no matter what time of day. And to be honest, the same is true for the bus. If you only took the bus in this town, you would think it is much more racially diverse than it is, although there is a great deal of diversity, it’s still a predominantly white community. If you take the bus then it’s vastly different. You’d also think that women outnumber men about 2 to 1.

\subsection{Reflection, Curiosity, and Exploration}
By far the most common theme we heard from participants was that Journeys \& Notes made them more reflective in different ways. For example, one person felt having the transit icons so closely associated with his identity made him question his transportation choices: ``I think I would try to bike more instead of drive. That was the powerful thing for me with the app since I didn't have any other contact [with others]. It really forced me to consider how I travel and if it was really necessary. I really liked it for that reason." Another participant felt that getting into the habit of writing while on a journey made him more aware of his commute and the world around him: ``I can only use my experience, but I think anyone who wants to actively be more aware of their commute would enjoy this. It allows reflection on a part of the day that is often dismissed as a necessary evil best ignored." Similarly, another felt that, ``Writers and thinkers would love it. It's a great way to be expressive. It might clue you in on things you had no idea about." This is particularly encouraging feedback in support of the journaling scenario for non-place based social applications. 

Another participant felt the app piqued her curiosity: ``the app just made me curious. It made me want to explore and look for people in the area. I really wanted to have a conversation with someone!" Similarly, someone else said: ``When I did get to read other's posts I felt some sense of connection --- like hey, other people are going through here and are sharing their experience. I liked the opportunity to pick up little secrets about places." She continued: ``Each time I logged in, I felt thoughtful and more conscious of my surroundings (except for having my head in my phone). But when I wasn't looking down, I was more aware of my surroundings because I wanted to write descriptively."

In response to seeing the haikus, another participant said: ``I did notice the Haikus! I liked them. They seemed to inspire me more than anything. They build up this idea that just around the corner I might meet someone unexpected."

\subsection{Public Transport}
We heard from several participants that they felt the app was most useful for public transportation (and not driving, or walking, or biking). For example, one woman said: ``I think the app is most useful for public transport, didn't log any trips I took by car or foot." This was echoed by several others. If we were designing the app from scratch, we would reconsider the decision to make it a general purpose transit app. We will expand more on this this feedback in the discussion section.

\subsection{Limitations}

Of course, not all participants enjoyed Journeys \& Notes. Some were just too busy with other things they valued more to give it much time: ``In transit, especially taking transit, I had other things I needed to pay attention to like not missing my stop and being alert to what was going on around me. I also don't like having my face buried in my phone on the bus because I think people do it to relieve social anxiety or out of boredom. I like being able to people watch- or gasp- even strike up a non-digitized exchange with someone."

Others just didn't feel connected to what it was trying to achieve, and felt systems like Facebook are more complete: ``At 53, I found nothing too interesting in it. You have Facebook and a few other things out there, so people check in there, though not me. I can see the benefit for entertainment of strangers commenting on stuff written by me, but since I was just outside Seattle, I saw no comments from anyone."

Another thought it was too buggy to enjoy: ``Why would anyone use it if this is what it is? What does it add to a trip? It is fiddly in alpha and doesn't look up addresses, and is yet populated with a lot of user content, so no, I'd not use it as it lives now."

%Not sure who, but then again, social media platforms operate with users who are less aware of privacy or humility as a virtue.

\subsection{Feature Requests}

One of the features that we had planned to add, but were not able to finish before these studies, was a statistics and tracking feature about journeys (e.g. quantified-self). Interestingly, this was something that was specifically requested: ``Also, I think a cool feature to add would be distance and time measurements. Maybe the app keeps track of how long the journey takes the user, or the user can add that information in. This would give others on the same trip more information hence why they use the app in the first place! Also adding a distance counter could be cool and easy, especially since the locations are already in the database."
This feedback motivated us to complete this exact feature in time for our release in the app store.

%\begin{table}[t]
%\centering
%\begin{tabular}{llrr}
%\toprule
% \textbf{Origin} & \textbf{Destination} & \textbf{Count} & \textbf{Percent} \\
% \midrule
% City & City & 73 & 2.1\% \\
% Home & Home & 39 & 1.1\% \\
% Neighborhood & Neighborhood & 27 & 0.8\% \\
% Train Station & Train Station & 23 & 0.7\% \\
% City & Home & 21 & 0.6\% \\
% Home & City & 20 & 0.6\% \\
% Neighborhood & City & 17 & 0.5\% \\
% Home & Building & 16 & 0.5\% \\
% Home & Office & 16 & 0.5\% \\
% Road & Road & 15 & 0.4\%
% \bottomrule
%\end{tabular}
%\caption{The ten most common (origin,destination) venue category pairs in journeys observed in the field %deployment.}
%\label{tab:origin_dest_pairs}
%\end{table}

%\begin{table}[ht!]
%\centering
%\scriptsize
%\begin{tabular}{lrr}
%\toprule
%\textbf{Mode} & \textbf{Count} & \textbf{Percent} \\
%\midrule
%Car & 4468 & 34.6 \\
%Airplane & 2200 & 17.0 \\
%Train & 1802 & 14.0 \\
%Bus & 1532 & 11.9 \\
%Foot & 947 & 7.3 \\
%Rocket Ship & 363 & 2.8 \\
%Bicycle & 321 & 2.5 \\
%Motorcycle & 263 & 2.0 \\
%Scooter & 183 & 1.4 \\
%Taxi & 159 & 1.2 \\
%\bottomrule
%\end{tabular}
%\end{table}

\section{Large Scale Field Deployment}

The app was publicly released in the Google Play store on October 22, 2014. Between then and September 20, 2015, there were 20,323 installations resulting in 23,697 user registrations (we purposefully do not maintain user identity across re-installs, resulting in duplicate registrations). During this time 9,435 people performed a total of 12,904 journey check-ins on 11,337 unique journeys. While most of these users performed just one journey check-in, there were 2,121 people with more than one check-in during this period, some with considerably more, with a maximum of 50 check-ins by one user observed. As one would expect, most journeys were relatively short with a long tail of very long journeys: the median distance was 54 kilometers (33 miles) and the mean distance was 1296 kilometers (805 miles).

%To get a better sense of the kinds of journeys people made during the field deployment, we looked into how the distribution of journeys broke down by the origin and destination venue categories. Table \ref{tab:origin_dest_pairs} shows the ten most common observed origin/destination pairs during the field deployment. 

We can also see how these journeys are distributed across different modes of transit (Figure~\ref{fig:modes} Right). While most people embarked on journeys on traditional modes of transit (Car, Airplane, Bus, Train, and Walking account for 84\% of check-ins), people did engage with our more playful and atypical modes of transit. We observed 58 people take journeys by skateboard, 125 people checked-in with a horse, 21 travelled by wheelchair, and 363 by the fantastical rocket-ship.

\subsection{Analysis of Contributed Notes}

Along with those check-ins, 1,989 users wrote 2,777 notes on their journeys, and 203 replies to existing notes. As one might expect, most of these notes were left in the default, {Notes \& Visitors} section (2,300 or 90\%). Some users also ventured from the default section and left their posts in {Tips \& Tricks} (122 or 5\%), {Love \& Hate} (51 or 2\%), {Secrets \& Stories} (45 or 2\%), or {Missed Connections} (16 or 1\%).

To explore how these 2,777 posts contributed to the ``sense of place" of the journey, we asked 5 annotators to categorize each post along five dimensions: (1) is the post poetic? (2) is it factual? (3) is it personal or intimate? (4) is it about memories of a past journey? and (5) is it about the present moment in this journey? We additionally wanted to identify posts that expressed the essence of the four personas that were critical to our design. We asked the annotators to also make judgements about whether the post author was (6) an everyday commuter, (7) a frequent flyer, (8) a quantified traveller, or (9) a journaling traveller. In this way, annotators labeled the data across 9 categories, and they could tag each post with any, all, or none of these categories. Table \ref{tab:annotated_posts} shows a summary of number of post tagged in each category, including Fliess' Kappa scores for the annotations. Fliess' $\kappa$ scores are computed, and in some cases indicate low agreement, perhaps reflecting the subjectivity of the task. We show approximate counts of each category type, by counting when at least two annotators agree on a labelling. Most posts were about the current journey (816), and the most exemplified scenarios were the Everyday Commuter (443) and the Journaling Traveller (373).  

\begin{table}[t]
\centering
\small
\begin{tabular}{lrr}
\toprule
\textbf{Category} & $\kappa$ & \textbf{Number of posts} \\
\midrule
Poetic & 0.41 & 77  \\
Factual & 0.22 & 536  \\
Intimate & 0.01 & 15  \\
About Past Journey & 0.06 & 46 \\
About Current Journey & 0.16 & 816  \\
\midrule
Everyday Commuter & 0.21 & 443  \\
Frequent Flyer & 0.09 & 76 \\
Quantified Traveller & 0.02 & 25 \\
Journaling Traveller & 0.06 & 373 \\
\bottomrule
\end{tabular}
\caption{A summary of post labels given by five annotators, where $\kappa$ is the Fleiss' measure of inter-rater reliability, and the number of posts counts  when at least two annotators agree.}
\label{tab:annotated_posts}
\end{table}

Below are some examples of notes written by users that exemplify the scenarios above. We only show posts from each category if at least two of the four annotators agreed the post belonged in that category.

\subsubsection{Poetic}

``From the shelter of my bus stop I watched the mist hug the walls of the building. Everything seemed to be wrapped in drops."

% ``I leave one train, only to get upon another. Exactly the same."

``The train is empty. Seats creak. The guard approaches silently. Gliding like he has done for years. I have no ticket to show him. I am not here."

\subsubsection{Factual}

``Grab a taxi or the BTS to the Victory Monument. From there head to the north-east side of the roundabout (you'll see all the minbuses) and speak to the women there. It should cost 60 Baht to the centre of Ayutthaya."

%\item ``Take a diversion via Brecon on the A470... Great change from heads of the valley."

``During the weekday, is easier to get coach ticket from the terminal bersepadu selatan (TBS), southern bound bus terminal."

% ``Find and play the piano at the airport. Relaxing."

\subsubsection{Intimate}

``I'm going to meet the love of my life for the first time, and if it doesn't work out I doubt I'll ever have the strength or will to overcome it. This is, regardless, a day of incredible happiness and growth. Optimism, more than anything."

\subsubsection{About a Past Journey}

``It is my second time back to TO after leaving it in 2005. This time marks as a special trip in memory of my one decade since graduated from U of T."

%\item ``Did this travel in 2007 and didn't regret it."

``My journey from the Philippines to the USA was years ago.  I migrated here with my family with hopes of a better future and life."

\subsubsection{About the Current Journey}

%\textbf{About the Current Journey:} ``The coffee is cheap and plentiful, and I'm on my way to Mexican food."
``Hi travelers, I'm a trucker and giving this app a try, headed to Ohio!"

%\item ``Aloha, Seattle! Off to BI for some ice cream."

``Got here just as the leasing office closed :-( Better luck tomorrow I hope but for right now, I'm getting tacos."

\subsubsection{Everyday Commuter}

``Just another ride to work at the shack. One day soon this ride won't be necessary and instead go on to do bigger things. That day will come soon I hope."

``My commute is farther from home than most people in history travelled in their entire lives."

%{\sc Everyday Commuter:} ``This my everyday, five day's out of the week route to my job at \{redacted\} Hospital."

\subsubsection{Frequent Flyer}
``I feel elated. Airborne. Rising higher. Soaring like a bird. With massive metal wings. And lots of creatures inside. People. Inside a bird. Odd."

``Left FAT airport at 2pm. It's never a long line so it took only 45 minutes to get in and in line for the plane. File out to SFO.  Got there in 45 minutes."
%\item ``Kiruna is small airport with nothing else than the essential. Souvenirs and restaurant."

\subsubsection{Journaling Traveller}
``Loved Prague.. could see living here with learning just a few words of Czech.  On to Poland!"

``Beautiful ambience, exotic wildlife, a calm weekend... What else do you want? The Boatbill strongly recommends this place."

``Stopped by Carkeek Park to see the salmon."
%\end{itemize}

%\end{itemize}

%Note that we do not find any notes from ``quantified travellers" worth sharing, as one would not expect these users to focus on note writing. We did find several users who logged their journeys religiously. 

\section{Discussion}

We believe we have identified a rich new space, ripe for future HCI research: developing mobile and social computing systems for engaging people in non-places. To explore this domain, we created Journeys \& Notes, a check-in app that augments journeys in the physical world with online places where people interact with other past, present, and future travellers of the same their journey. Our intention was to design a system that might surface new experiences and address unmet needs of travellers passing through non-places. We grounded our design decisions around four archetypal scenarios where non-place travellers might encounter social computing: the everyday commuter, the frequent flyer, the journaling traveller, and the quantified traveller.

Through two small scale user studies, and one field deployment, our preliminary investigation shows evidence that social computing can indeed imbue a transient physical space with basic characteristics of a place: elements of a history, a culture, a community, and mechanisms to foster a heightened sociality among those that pass through it.

Our small scale interview studies revealed a broad range of reactions, both positive and negative, to individual features of the design. They also present strong evidence in support of the central premise of this work, that there is opportunity for social computing to grow online communities that are rooted at non-places. The most valuable aspect of Journeys \& Notes that emerged from these interviews seemed to be the fact that through a small set of very basic interactions (check-in mechanisms, and note compositions), the system could support a range of different uses: self-reflection, atemporal interactions with others, discovery of new places, self-tracking and community building.

Although the app-store release of Journeys \& Notes has yet to successfully generate sustained usage or foster dense, anonymous, hyper-local communication channels (see the limitations section below), it has nevertheless been successful at providing a window into the needs and thoughts of the non-place traveller. Because communication in Journeys \& Notes is atemporal, the notes that people write are like messages in bottles tossed into the sea, composed for future travellers of the journey to read, regardless of whether anyone ever finds them. The content of the messages examples above give us a fascinating glimpse into the range of different things people post about, and the potential place-making impacts such a communication channel might have in transient spaces. We take these messages as further evidence that Journeys \& Notes speaks to interactions and scenarios that resonate with people, and yet are not fully explored by existing research or technologies. The richness and opportunity for social computing for non-places is clear.

\subsection{Limitations and Lessons Learned}

In our evaluations of the design, we uncovered a number of limitations and lessons that might be valuable to future researchers in this area. Many of these lessons stem from the tensions between being a low fidelity research project, while also being a product available in an app market. On the surface Journeys \& Notes appears somewhat polished, but as an experimental app, it lacked many features and details that product teams would typically invest in.

In order to truly get sustained usage of the app, we would have needed to invest more into cracking the two-sided problem of geographic sparsity hindering the discoverability of posts. A number of approaches might have helped. Instead of a world-wide release in an app store, we could have embraced geographic-contagion model for the release, focusing on one dense geographic region to start and scaling out. We also would have needed to embraced device notifications as a mechanism to inform people when other activity happens near them, or on the posts that they leave behind, thus increasing interactions, and discoverability of people and content.

We also found that post-sections (e.g. missed-connections, tips \& tricks, etc.) were barely used at all, likely because of a flawed UI design that made them difficult to discover. While we included these as a way to nudge users to focus their conversation around fixed topics, in retrospect, people seemed post compelling and interesting content without using them, so they may not be necessary after all. 

There were a number of features that we explicitly chose not to support that ended up being requested by users. We wanted to create an ephemeral experience similar to other online communities \cite{bernstein20114chan}, and designed our posts so that they were only visible when people were actually checked-in to their journeys. People didn't like this. Here's an example review written by one of our users: ``i want to be able to delete a journey and review the notes I've made\ldots bc right now, i have to log in another journey to see my notes and it looks so messy now." We also wanted to design a text only experience, but users today are so used to rich media, that there was a strong desire to see photos in posts.

%We believe the system reduces friction of participation by not requiring authentication, especially not requiring a Facebook login; however, in our attempt to make it easy for people to participate and with automatically assigned pseudonyms, we made it difficult for people to change their name, as one of the app store reviewers mentions: ``I like the random username generation, but I wish it could be re-randomized for different options." This also hinders growth by not leveraging existing networks, as some users had the expectation that they could import their existing friends from other services. Balancing anonymity and 

Some people, especially the quantified travellers, found it too much work to have to manually enter every journey. While we made it easy for people to check-in to previous journeys by simply clicking on them; other apps make it even easier. For example, the Moves app uses the geolocation features in smartphones to automatically log people's activities \cite{cutler_2013}. While this auto logging opens up challenges for privacy and device battery-life, it is a feature some people expect to have. Future iterations should consider giving people this option.   

The system tried to capture all types of a journey's modalities, from bus to plane to skateboard. This reduces the possibility to optimize for one type of community; in retrospect, the system could have been created with one specific community in mind, similar to what Waze has done by focusing on automobiles. This need to focus on a specific community was highlighted by one of the emails we received from one of our most avid users: ``I think that you should work on a trucker version. Drivers could leave notes about truck stops, warehouses, road hazards, places with truck parking and such. I think that it would be a hit. (\ldots) I am a truck driver. I have been driving for about 20 years."

\subsection{Future Research Opportunities}

This work opens the door to a number of different promising areas of future research. By strengthening community structures around common commute paths, non-place communities could help finally realize the original vision of peer-to-peer ride-sharing (not the ``ride-for-hire" form of ride sharing popularized by Uber and Lyft). Beyond specific end-user functionality, we also envisioned these type of systems being useful sources of civic data that communities can use to better understand mobility patterns, transit choices, and people's perceptions of their surrounding environments. We also believe that developing technologies specifically for non-places can help  meet the challenges of rebuilding and strengthening communities in the 21st century. The increased occupation of non-places, and the growing communities these non-places contain, makes them increasingly central to social development and interaction, and properly designed technologies can transform Milgram's familiar strangers into knowledge sharing and human contact even in an asynchronous and pseudonymous fashion. %Technology can transform non-places into the focal points for modern communities, even as—--perhaps especially because—--communities themselves are becoming more flexible and frequently shifting.

%\section{Conclusions}

%- identified a rich, novel space for future research.
%- provided a proof of concept design illustrating that there are unmet needs for augmenting non-places with social computing.
%- early evidence suggesting that social computing can impact peoples experiences at physical non-places, perhaps even augmenting them with a virtual sense-of-place.
%- types of content people leave suggests
%- while much more work is needed to make a production quality system, we have opened several new doors with this work to encourage future work, both in industry and academia.

%\section{Acknowledgements} 
% Thanks to odesk content writers (list names)

% REFERENCES FORMAT
% References must be the same font size as other body text.
\newpage
\balance
\bibliographystyle{SIGCHI-Reference-Format}
\bibliography{j_and_n}

\end{document}